# A Work-Centered Approach for Cyber-Physical-Social System Design: Applications in Aerospace Industrial Inspection

Garrick Cabour
Ecole Polytechnique de Montreal
Department of Mathematics and Industrial Engineering
2500 Chemin de Polytechnique, Montréal, Québec H3T 1J4, Canada
Email: Garrick.cabour@polymtl.ca

Elise Ledoux
Universite du Quebec a Montreal
Physical Activity Department
2098 Rue Kimberley, Montréal, Québec H3C 3P8, Canada
Email: ledoux.elise@uqam.ca

Samuel Bassetto
Ecole Polytechnique de Montreal, Canada
Department of Mathematics and Industrial Engineering
2500 Chemin de Polytechnique, Montréal, Québec H3T 1J4, Canada
Email: samuel-jean.bassetto@polymtl.ca

## Abstract

Industrial inspection automation in aerospace presents numerous challenges due to the dynamic, information-rich and regulated aspects of the domain. To diagnose the condition of an aircraft component, expert inspectors rely on a significant amount of procedural and tacit knowledge (*know-how*). As systems capabilities do not match high-level human cognitive functions, the role of humans in future automated work systems will remain important. A Cyber-Physical-Social System (CPSS) is a suitable solution that envisions humans and agents in a joint activity to enhance cognitive/computational capabilities and produce better outcomes. This paper investigates how a work-centered approach can support and guide the engineering process of a CPSS with an industrial use case. We present a robust methodology that combines fieldwork inquiries and model-based engineering to elicit and formalize rich mental models into exploitable design patterns. Our results exhibit how inspectors process and apply knowledge to diagnose the component's condition, how they deal with the institution's rules and operational constraints (norms, safety policies, standard operating procedures). We suggest how these patterns can be incorporated in software modules or can conceptualize Human-Agent Teaming requirements. We argue that this framework can corroborate the right fit between a system's technical and ecological validity (system fit with operating context) that enhances data reliability, productivity-related factors and system acceptance by end-users.

# I. Introduction

In the aerospace domain, aftermarket services of engine components (maintenance, repair, and overhaul or MRO) account for more than 50% of aircraft manufacturers' revenues (Cohen et al., 2006; CRIAQ, s. d.; Muñoz & Morwood, 2020; Rolls-Royce Holdings plc, 2019). Aftermarket servicing is articulated around the expertise of trained operators, cutting-edge technologies and highly standardized processes (Johnston, 2017). At the center of MRO, industrial inspection is a manual-extensive and bureaucratic labor in which inspectors scrutinize aircraft parts for serviceability based on standard procedures (C G Drury, 1999; See, 2012). In view of the ever-increasing customer requirements (quality, cost, delivery); and increased competitiveness through the development of cutting-edge technologies, this human-specific task has benefited from significant automation intentions in the past three decades (Muñoz & Morwood, 2020; See, 2012; See et al., 2017). However, it still presents numerous challenges in aerospace due to the dynamic, information-rich and regulated aspect of this safety-critical domain.

High-level human cognitive functions - such as sense-making and decision-making – are partially mastered by machine intelligence (Abbass, 2019; Dellermann, Ebel, et al., 2019). In spite of recent advances in machine learning algorithms, current systems do not grasp large problem space within real-work settings (i.e., set of explicit and fuzzy rules) (Dhuieb et al., 2015; Zheng et al., 2017). Also, machines behave according to algorithms for anticipated situations (Zheng et al., 2017). However, the context of industrial operations is represented by uncertainty, multifaceted problem space and heterogeneity of information (Belkadi et al., 2019; Miller et al., 2016). Experts in industrial processes rely on intuitive methods (case-based reasoning) and adaptative control when dealing with unforeseen situations that *leverage constraints on problems and solutions in intelligent ways* (Flach et al., 2017). Despite their problem-solving skills, human cognition is limited. The rate of human error in visual inspection ranges from 20% to 30% (See, 2012). New generation sensors can enhance perceptive capabilities of humans in order to make more informed decisions in automation-assisted inspection scenarios (Agnisarman et al., 2019).

Combining Human and Artificial Intelligence is a promising direction to enhance process efficiency and meet customer requirements. A Cyber-Physical-Social System (CPSS) is a suitable solution that adds a Human-Agent Teaming (HAT) layer in the design of intelligent systems (Xiong et al., 2015). It envisions humans and agents in a joint activity to produce better outcomes than each of the two could produce separately (Jiao et al., 2020; G. Klein et al., 2004; Rahwan et al., 2019). In particular, it is a matter of combining intuitive reasoning and contextual knowledge application from humans (*soft data,* often machine-unreadable) with an agent's computational power and analytical skills (Belkadi et al., 2019; Dellermann, Calma, et al., 2019). Several HAT frameworks were developed for "physical" tasks (e.g., *pick and place* and/or in controlled environments) (Mateus et al., 2019; Wang et al., 2019). However, HAT for industrial inspection is converging towards a more cognitive level. A combination of cognitive and computational skills are required to face a large problem space composed of many variables, fuzzy rules and data heterogeneity (Abbass, 2019; Jiang et al., 2004). Therefore, there is a growing need to elaborate HAT design methods that encompass collaborative cognition between humans and agents in CPSS (IBM, 2018; Jiao et al., 2020).

**Research Gaps 1 & 2** – Limited machine skills in processing real-world variables, fuzzy rules and *soft data*. Limited human cognitive skills. Limited methods for Human-Agent Teamwork conceptualisation in cognitive tasks for real-world problem space.

Industrial inspection in aerospace (MRO) is characterized by a vast problem space where operational demands and institutions rules generate a complex workflow (G. A. Boy, 2017; T. L. Johnson et al., 2019; See, 2012). Erroneous decisions can lead to human fatalities and loss of costly components in this safety-critical domain (See et al., 2017). Diagnosing a part requires the integration and processing of several sources of knowledge, sometimes contradictory, from the problem space (Agnisarman et al., 2019; Flach et al., 2017). The overall success of the process depends heavily on inspectors' expertise, to the extent that the knowledge deployed is up to 50% tacit (know-how*)* [i.e., does not appear in standard operating procedures (SOPs)] (T. L. Johnson et al., 2019). A work-centered approach can bring valuable benefits to CPSS design by formalizing the tacit and situated dimensions of industrial inspection.

Capturing and reusing operational knowledge is key for the design and implementation of future intelligent systems (Belkadi et al., 2019; T. L. Johnson et al., 2019; Roth et al., 2019). Based on fieldwork inquiries, a work-centered approach elicits how operators integrate knowledge during decision-making, deal with institutions (rules, policies, norms), and interact within the socio-technical system. Data gathered from fieldwork inquiries must be aligned with technical development in order to be exploitable and address engineering challenges (Feigh et al., 2018; Robert R. Hoffman & Klein, 2017). Recommendations provided by fieldwork researchers (e.g., human factors practitioners), while extremely valuable in conferring ecological validity[1] to the system being designed, are seldom actionable (Emmenegger & Norman, 2019). Often, data are conveyed to designers in a purely descriptive form, lacking integrability and guidance for design engineering milestones (R.R. Hoffman & Deal, 2008; M. Johnson et al., 2014). To outperform those limits, some researchers adopted a model-based approach using input from fieldwork data to inform software specification (Dhukaram & Baber, 2016), safety assessment (Vries & Bligård, 2019), agent-based modeling (Elsawah et al., 2015) or knowledge management (T. L. Johnson et al., 2019). However*,* how

---

[1] Ecological Validity: fit between system capabilities, work requirements, operating context and end-users needs (Cabitza & Zeitoun, 2019)

these models are used to solve design challenges remains unsettled. Further research is required to bridge this gap.

**Research Gap 3 & 4** – Lack of methods to elicit and translate fieldwork data into relevant tangible design patterns for CPPS features specifications and ecological validity assessment (software module, HAT requirements).

This paper proposes a work-centred approach for CPSS design that advocates a systemic analysis of industrial inspection work to understand the *nature of work being transformed and what makes it challenging* (Roth et al., 2019). The common thread is to capture and reuse operational knowledge to guide technical development, system's implementation and HAT requirements. We investigate how fieldwork data could be formalized into actionable input for software engineering and support the conceptualisation of a CPSS in real-world settings. This conceptualization ultimately explores ecological validity early in the engineering cycle to ensure the CPSS will meet operational requirements and process all variables within the problem space. Specifically, we seek to answer the following research questions:

- **RQ1**: How inspectors inspect and sentence[2] a part *in situ*? What are the variables, rules and constraints that shape their cognition (decision-making and sense-making process) and actions? What are their interdependencies with the other actors of the socio-technical system during sentencing?
- **RQ2**: What type of fieldwork data should be transferred to software and automation engineers? How should the fieldwork data collected be formalized into actionable design patterns that align with the conceptualisation and preliminary design of a CPSS?

The remainder of the paper is organized as follows. Section II provides an overview of the industrial context in which the project takes place. Section III clarifies the theoretical background and related research, from an analysis of literature related to industrial inspection, socio-technical systems and empirical models. Section IV exposes the research method related to this study, specifying the building steps of the models from fieldwork data. Section V presents a case study in aircraft maintenance inspection. We exhibit the different models and fieldwork results. Section VI summarizes the results of the study and its outcomes, as well as the further implications of the approach for intelligent system design.

---

[2] Sentencing = decision about their status (acceptable as is, salvageable, unserviceable) and prescription of corrective actions (defect removal)

## II.  The SARA Project

Increased automation is expected to enhance production-related capabilities, reinforce safety, and minimize cost and delivery time. SARA stands for *Système d'Analyse et de Réparation Automatisée* (Automated Visual Inspection, Sentencing & Dressing). The project involves a consortium of academics, R&D institutes and industrials that aims to automate the sentencing of complex service-run components in the aerospace sector. *To the consortium's knowledge, no such system is currently available or under development* (CRIAQ, s.d.). Three components out of a thousand are candidates for the project (turbine disks, high-pressure turbine shaft and fan blades). They are chosen for their representativeness of the plant's fleet of parts, ranging from complex geometries to open flat surfaces. *Part of the challenge is to automate the data from the inspection into machine-readable criteria for the sentencing* (CRIAQ, s.d.). Indeed, inspectors rely on rich mental models when facing unforeseen situations but each *machine requires programming specific to each situation* (CRIAQ, s.d.). The other challenge concerns Human-Agent Teaming (HAT) in complex work-settings. Since SARA is expected to perform some tasks that were previously carried out by inspectors, in cooperation with them, the conceptualization of the human-agent interaction should be addressed early in the engineering cycle. It is essential to capture how Work-Is-Done (WAD): how inspectors decide (rules, variables and constraints) and deal with the institution's rules (procedures, norms, safety policies), and also how these factors shape their cognition and action. Therefore, our contribution to the SARA project is to first elicit and formalize contextual information based on the work of subject-matter experts (e.g., defect detection cues, sense-making and decision-making). Then, we translate this data into a set of actionable design patterns that support technical development, therefore guiding the conceptualisation and assessment of HAT to facilitate system's design and implementation.

## III.  Background and Related Research

This section presents a selection of papers, best practices and standards that are relevant to the challenges being addressed. First, we present a state-of-the-art of relevant knowledge about inspection, whether it be manual, automated or semi-automated. Second, we review the most pertinent methods to collect and model fieldwork data intended for system design. Finally, the concept of Human-Agent Teaming will be presented with the existing frameworks.

### III.1.  Manual, Automated and Automation-Assisted Inspection

Industrial Inspection is a crucial step in any manufacturing process that aims to certify the quality of production (Agnisarman et al., 2019; Baudet et al., 2013; Kujawińska & Vogt, 2015). Standard procedures issue quality norms, types of surface anomalies expected (defects) and acceptance/rejection criteria (Colin G. Drury & Dempsey, 2012). Inspectors are responsible for assessing deviations from the defined standards by performing multisensory analysis on products (vision, proprioception, touch) (Colin G Drury & Watson, 2002; T. L. Johnson et al., 2019). Inspection is a highly cognitive process that can be grouped into 5 stages according to several authors. *Visual search* and *decision-making* are the most complex and error-prone (Table 1) (Colin G Drury & Watson, 2002; Jiang et al., 2004; See, 2012). Both stages incorporate a substantial amount of knowledge (procedural, explicit, tacit) to interpret and apply, where inspectors' expertise determines the final performance of the process (Colin G. Drury & Dempsey, 2012; See et al., 2017). While inspection seems procedural and linear, Drury defines it as an *"ill-structured work because there is no simple step-by-step procedure which will ensure success, and because there is usually no knowledge of task success available during the task"* (C G Drury, 1999, p.3). In other words, the process is highly situated, involving dynamic decisions[3] in which the current and desired state of the situation is not completely defined. Two studies also concluded that inspectors are rational decision-makers who use economic variables such as '*the likelihood that an object will be defective*' to assess the need to continue the inspection process (Benjamin et al., 2009; Jiang et al., 2004). Thereby, it seems relevant to identify the various types of criteria (variables) that shape inspectors' decision-making process (whether they are economic, technical or social). In our knowledge, empirical research on industrial inspection documented the practice from a person-oriented perspective. However, we support the need to extend the scope to a socio-technical perspective, to see how inspectors are integrated and interrelated to several components of a work system (actors and agents). This macro perspective will strengthen the analysis of the potential automation's effect on the interdependencies between actors.

*Table 1- List of tasks and functions required to perform an inspection (adapted from Wand and Drury,1989; Jiang and coll., 2004; See et al., 2017)*

| | Task | | Sub-Task Description | Type of skill | Cognitive functions required |
|---|---|---|---|---|---|
| 1. | **Set up** | 1.1 | Routing inspection equipment, aids and parts | Manual and cognitive | Memory |
| 2. | **Present** | 2.1 | Orient the item | Manual | - |
| 3. | **Search** | 3.1 | Search the item (visual, tactile and proprioceptive) | Cognitive | Attention, perception, memory |
| | | 3.2 | Detect the flaws | Cognitive | Detection, recognition, memory |
| 4. | **Decide** | 4.1 | Classify the flaws | Cognitive | Recognition, classification, memory |
| | | 4.2 | Decide about the item, comparison against quality standards | Cognitive | Judgement, classification, memory, sense-making |
| 5. | **Respond** | 5.1 | Dispatch the item | Manual | - |
| | | 5.2 | Record item's information | Manual and cognitive | Memory |

Automated inspection systems can provide solutions to increase the efficiency, safety and speed of the process (See et al., 2017). Their introduction is generally motivated by the desire to address the shortcomings of traditional inspection: speed-precision trade-off, expensiveness, human error, time required to train qualified inspectors (Agnisarman et al., 2019; Jiang et al., 2004). The combination of new generation sensors and image processing techniques enhances data quality & processing, traceability, reduces uncertainty and enables multi-layer analysis in real-time (Ferraz G. T. et al., 2016; S. Jordan et al., 2018). Automated inspection systems can detect up to 98% of defects with a false alarm rate of about 2% (See, 2012). Despite these achievements, Agnisarman et al. (2019) present some shortcomings of automated inspection systems:

1. Inspections systems support the *Search* phase of inspection (task 3, Table 1) and the more advanced ones are also able to classify defects (sub-task 4.1, Table 1). None of the reviewed systems support the decision phase. During this task, inspectors *process and make-sense of data from multiple sources* in order to generates correct outcomes (Agnisarman et al., 2019). In aircraft maintenance, this involves interpreting the convergence of a defect with associated rules (procedural and tacit), safety policies and other local/global protocols (multi-criteria decision-making).
2. The performance of these lab-developed systems needs to be validated through deployment in a real context to appraise its ecological validity (i.e., the fit between technologies' features, taskwork requirements, operational and user needs) (Agnisarman et al., 2019; G. A. Boy, 2020). Discrepancies between technical and ecological validity generally hamper productivity-related and social aspects (e.g., user acceptance and perceived usefulness) (Cabitza & Zeitoun, 2019).
3. Partial automation of inspection implies the subordination of the system to the operators (Agnisarman et al., 2019; Jiang et al., 2004). The system is expected to execute some tasks that were previously carried out by inspectors and new ones will emerge from the Human-Agent interaction (G. A. Boy, 2020; T. L. Johnson et al., 2019). Therefore, the quality of the Human(s)-Agent Teaming (or HAT) must be conceptualised, designed and tested on real use-case scenarios (Cabitza & Zeitoun, 2019). This point is individually developed in Section III.3.

---

[3] Dynamic decision: interdependent spiral of decisions that influence each other (Brehmer, 1992).

These limitations emphasize the need to first carry out *in situ* analyses in order to understand the physical, cognitive and social dimensions of the inspector's work activity in real-world settings (Work-As-Done or WAD). Based on this understanding, the Human-Agent Teamwork must be conceptualised during the system design cycle to ensure ecological validity. What follows is a brief review of the fieldwork methods best suited for the systemic analysis of work activity in a situated context (WAD), and the formalization of fieldwork data through empirical models.

### III.2. Fieldwork Inquiries, Data Elicitation and Empirical Modeling

Fieldwork inquiries allow for the study of people in their real-work environment (G. Boy, 2016). They aim to examine a work system through a set of bottom-up methods: observation, interviews, focus groups (St-Vincent et al., 2014). Data collection and interpretation is usually performed from the actor's point of view [i.e. the subject-matter experts (SME)]. The *naturalistic decision-making* paradigm emphasizes how the cognition of experts is solicited in complex work-settings (Gary Klein, 2015). Cognitive Task Analysis is an analytical method that examines and decomposes the mental models of experts in their work context (Gary Klein, 2015). Activity analysis is a systemic approach to work situations that focuses on how work is done to produce the predetermined results from a holistic perspective that includes physical, mental and social aspects of human work (St-Vincent et al., 2014). The method considers the domain constraints (determinants) as modulating factors of operators' cognition and action (e.g., problem-solving strategies implemented by the actors to overcome procedures' limitations) (St-Vincent et al., 2014). These two methods are valuable for understanding the socio-cognitive aspects of inspectors' work activity (decision-making, sense-making, collaboration with other actors) and for situating their role within the socio-technical system.

When describing a complex decision-making process, researchers should focus on formalizing the semantics of the problem space (Dellermann, Calma, et al., 2019; Simard et al., 2017). That is, identifying and describing the sequence; and the interrelated factors that compose the *decision flow*: domain knowledge (explicit and tacit), variables, constraints, rules, protocols and other actors that shape the cognition and action of subject-matter experts in situated context (Elsawah et al., 2015; Hobballah et al., 2018). Fieldwork researches in real-world settings have shown that "*robust decision-making depends on macro-cognitive phenomena at the meaning-level, the knowledge-level and the context-leve*l (Robert R. Hoffman & Klein, 2017). For industrial inspection, this involves the elicitation and representation of defect detection cues, sensemaking and decision-making process. The step after is showing how these elements are interrelated (Meinherz & Videira, 2018).

Fieldwork researchers (e.g., human factors practitioners) experience difficulty in integrating fieldwork data in engineering design (Baxter & Sommerville, 2011; Wilson & Sharples, 2015). Translating textual data into actionable design objects, such as a machine-readable program, is one of the challenges encountered (Elsawah et al., 2015; Scheller et al., 2019). Other obstacles concern the lack of guidance provided to designers with the patterns/data obtained (Emmenegger & Norman, 2019). Informing designers about work complexity is an important step. But a clear plan to implement an abstraction of data in designers' work packages must follow (M. Johnson et al., 2014). Finally, the models generated must accurately reflect fieldwork data (i.e., without being "lost in translation") (Dhukaram & Baber, 2016). A rigorous methodology for data analysis and translation is essential but only a few papers explain in detail how to proceed (Scheller et al., 2019). Thus, the methodological content of this paper will be presented with a deep level of granularity (see Section IV).

The model-based system engineering approach has continued to draw attention over the years with respect to cross-functional integration & collaboration (Madni & Sievers, 2018). Models abstract system complexity and promote communication between stakeholders and mutual learning (Edmonds et al., 2019; Madni & Sievers, 2018). They also formalize and endorse the integration of experts' knowledge into computer programs (Vries & Bligård, 2019). Dhukaram & Baber (2016) derived qualitative fieldwork data into UML models to design decision aids. Vries & Bligård (2019) developed 5 different systemic models to encourage designers adopting a sociotechnical vision of safety. Johnson and al. (2019) used a combination of task analysis and task decomposition to systemically disaggregate process tasks in terms of knowledge and skills required. To be efficient and avoid missuses, the modeling purpose must be explicitly specified (Edmonds et al., 2019). In this paper, a combination of descriptive and formative models will be used (Section IV). Descriptive models afford comprehension (i.e., "what is") whereas formative models afford prospective formation (assessment) and exploration (i.e., "what should be") (Le Coze, 2013).

### III.3. Pre-Conceptualisation of Human-Agent Teaming

The role of humans in future automated work systems will remain important and greater collaboration between humans and automation is expected (Pettersen, 2018). Achieving the full potential of technological development doesn't rely on maturing the artifact alone, but rather on maturing a symbiotic Concept of Operations (Abbass, 2019; G. A. Boy, 2020). An operational concept (re)allocates tasks, roles and responsibilities among individuals and agents (Pritchett et al., 2016). The socio-technical context must be considered in the design of CPSS where economics, safety and performance-related factors are evaluated (Pritchett et al., 2016). It also makes possible to *identify the constraints that place clear limits on what is possible with respect to which functions individuals and/or agents can take on* (Roth et al., 2019). Technological capabilities, safety policies, criticality of the outcome, explanation capacity of the machine may constrain the possible Concept of Operations. For a highly cognitive work such as inspection, Human-Agent Teamwork should reassign (and redefine) tasks, decision steps and decision variables. The work system needs to be reorganized based on an understanding of how work is currently done (WAD), work requirements and *associated challenges* (Roth et al., 2019). This prerequisite is generally neglected in function allocation methods and leads to the conceptualisation of inefficient, unsafe or unrealistic Concept of Operations (Roth et al., 2019). Stern & Becker (2019) introduced a framework to link a human-oriented approach to a Cyber-Physical System design with a focus on cognitive collaboration among workers and automation. They concluded on the need to explore practical tools to develop CPSS with an industrial use case study. There is an urge to develop a set of methodological tools to conceptualize Human-Agent Teamwork for complex socio-cognitive work in industries.

To summarize the gaps identified in this literature review: 1) Future production systems will incorporate workers and automated agents in a cognitive collaboration where there is a lack of conceptual frameworks and application in real-work settings; 2) HAT requires actionable patterns and guidance from fieldwork researchers (e.g., human factors practitioners or knowledge engineers); 3) Fieldwork data integration is difficult to orchestrate in systems design and requires methodological consolidation.

This paper contributes to advancing knowledge on the conceptualisation of HAT for real-work settings. More specifically for the first stage that consists of understanding, describing and formalizing WAD in an exploitable format for technical development and implementation of CPS(S). This understanding will further be aligned with technological capabilities, interdependencies of tasks and human-automation interaction requirements (Robert R. Hoffman & Klein, 2017; M. Johnson et al., 2014).

# IV. Methodology

## IV.1. Fieldwork (Empirical) Data Collection

The empirical basis of this paper originates from a case study in the Aircraft Maintenance, Repair & Overhaul (MRO) domain in Canada. The facility operates in-service aircraft engines and the workflow can be simplified as follows: on arrival at the factory, aircraft engines are first disassembled and then each component goes through a robust inspection process to certify it is free of defects. If not, damaged components are either repaired or replaced.

### IV.1.1. Internal Documentation Analysis (Work-As-Imagined – WAI)

Aircraft MRO is a standardized and normalized work environment where procedures ensure that components meet safety policies. As an internal documentation analysis, we reviewed Standard Operating Procedures (SOPs) and the reference manual. These documents contain all the instructions (steps, requirements, 3D drawings of the parts), explicit knowledge (dimensions, tolerances, limits) and resources needed (magnification system, directional lightening) to complete an inspection/sentencing task. The analysis focused only on inspection aids and procedures. By analogy, inspectors act as information processors: they compile data from their environment and compare it with information stored in their own memory and/or in instruction manuals (See, 2012). Therefore, capitalizing the instructions (and format) provided was necessary to understand how inspectors interpret them *in situ* [i.e., what strategies they implement (Work-As-Done - WAD) to compensate for the limitations of inspection aids whether tools or documentation (Work-As-Imagined – WAI)].

### IV.1.2. Fieldwork Inquiries (Work-As-Done – WAD)

Fieldwork data collection involved a mixture of on-site observations and experiments (*microworld*), *think-aloud* protocols and individual and group interviews (*focus group*). We elicited different types of knowledge (either procedural, conceptual, explicit or tacit) by varying our collection methods (Figure 1). Inspectors use up to 50% of tacit knowledge during their inspection process (T. L. Johnson et al., 2019). This type of knowledge stocked in the head of experts contains "*contextual information about how to best detect* [characterize] *and diagnose product defects*" (Johnson et al., 2019). Therefore, capturing and modeling fieldwork data are valuable to i) formalize the underlying "variables" shaping inspectors' decision-making process (i.e., rules, knowledge, constraints, protocols) ii) to provide actionable recommendations and inputs to automation engineers.

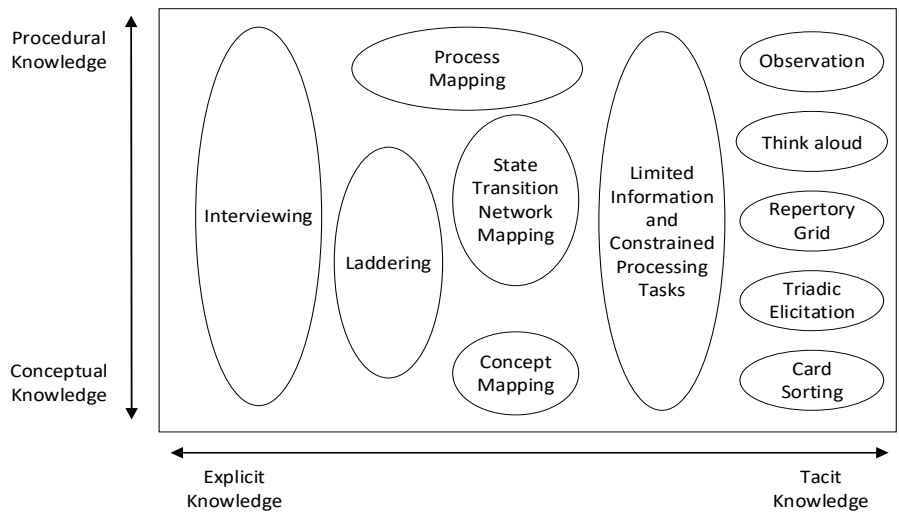

*Figure 1 - Comparison of knowledge elicitation methods according to the type of data targeted (Milton, 2003)*

Ten inspectors were interviewed individually using 3 types of interviews. General knowledge about inspection activity was first gathered with Semi-Structured Interviews (Adams, 2015). We then used Elicitation Interview (Hogan et al., 2016; Vermersch, 2010) and Self-Confrontation Interview (Theureau, 2010) methods to meet several objectives. First, to understand the cognitive processes shaping their actions/intentions. Then, to extract contextual knowledge and detailed descriptions of their activity after observation sessions. Semi-structured interviews provided explicit and general knowledge about inspection and the socio-technical system (Figure 1). We then gained tacit insights on inspectors' work activity with Elicitation and Self-Confrontation interviews. These interviews, supported by fieldwork data (textual, audio or video contents), allow inspectors to comment on specific points of their activity, for example the unconscious automatisms they have developed with expertise. They can be carried out immediately after observation or after processing, organisation and selection of fieldwork data to comment.

We conducted several types of on-site observations to understand how the entire sentencing process unfolds; from workpiece collection to the stamping of administrative paperwork (Table 2). We observed inspectors – with experience ranging from less than a year to more than ten years – working on two aircraft components selected for the SARA project: fan blades and high-pressure turbine shafts (HPT shaft). First, we "*shadowed"* inspectors in real work context (i.e., without interacting with them). This made it possible to gain insights into the "observable" part of work in an ecological perspective. The following elements were scrutinized: communication with other workers, nature of these communications (problem-solving?), physical operating methods (e.g., the inspector rotates the aircraft part under the light source), information intake operations and sources (computer, paperwork, aircraft parts, storage racks), direction of gaze, movement around the workplace.

*Table 2 - Fieldwork Inquiries Characteristics*

|  |  | Inspectors | | | Managers, Planners and Engineers | | | Material Review Board | | | Operators | | |
|---|---|---|---|---|---|---|---|---|---|---|---|---|---|
|  |  | n | Sessions | Minutes | n | Sessions | Minutes | n | Sessions | Minutes | n | Sessions | Minutes |
| Fieldwork Inquiries* | Observations | 7 | 11 | 1110 |  |  |  | 1 | 1 | 90 | 3 | 3 | 390 |
|  | Interviews | 11 | 26 | 810 | 7 | 5 | 275 | 2 | 2 | 60 | 4 | 5 | 60 |
|  | Experiments | 4 | 6 | 660 |  |  |  |  |  |  |  |  |  |
| Sub-Total |  | 22 | 43 | 2580 | 7 | 5 | 275 | 3 | 3 | 150 | 7 | 8 | 450 |
| Total |  | 39 | 59 | 3455 |  |  |  |  |  |  |  |  |  |

* The table excludes the time spent for internal documentation analyses (WAI), data extraction from the Enterprise Resource Planner (ERP) and informal stakeholder inquiries.

In order to delve deeper into the cognitive sphere of inspectors' work, we then performed focused observations. Since inspecting and diagnosing a part condition is highly cognitive, inspectors were asked to *Think Aloud* (i.e., to make *concurrent verbalization while performing a task*) (Güss, 2018). This protocol is extremely effective to *go beyond what is merely observable* (St-Vincent et al., 2014). Specifically, we asked the following questions for each action taken by the inspectors: "What are you thinking right now" or "What are your intent?" "What are you doing?" / "How will you do that?" "How did you come to this decision?" "Are there any alternatives to this course of action?" We focused on the decision-making flow of inspectors to determine how procedural and tacit knowledge were applied (Figure 1). The following elements were scrutinized: cognitive tasks (steps), information intake mechanisms, problem assessment and investigation of available courses of actions (sense-making), decision variables, evaluation of decision outcome, operational constraints, rules and strategies applied (expertise), deviations from SOPs.

*Think-aloud* protocols were also deployed in fieldwork experiments *(microworld)*. *Microworld* refers to the reproduction of a laboratory-based environment that represents the working context as authentically as possible. By doing so, we explore the problem-space and information-rich environment of inspectors' complex decision-making process with a deep level of granularity (Figure 2). Factory engineers asked two experienced inspectors to map 5 aircraft components (4 Fan Blades and 1 HPT Shaft). Mapping a component involves recording each detected defect with their relative metrics (dimensions in length, width and depth) and outcomes (accepted, repairable, unaccepted). Then, we asked three experienced inspectors to inspect each of the mapped components while thinking aloud, except for the HPT Shaft where we could only conduct the experiment with one person. Decision outcomes, variables and defects metrics were known beforehand, allowing for a richer simultaneous self-confrontation with inspectors throughout the process. We emulated different scenarios to explore the problem space from different angles (e.g., limitation of certain information during sentencing) (Figure 1). As we interacted, interrupted and modified the natural working conditions towards a laboratory-based environment, the *microworld* slightly decreased the contextual fidelity of real working conditions (ecological fidelity) but increased the granularity of data elicited (Figure 2).

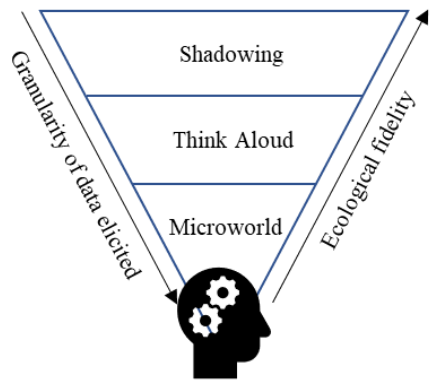

*Figure 2 - Relation between methods used, granularity of data elicited and ecological fidelity*

In addition to analysing the work of inspectors, we also interviewed/observed actors (n=8) that interact with them in the socio-technical work system: *Production Supervisors and Planners, Polishers, Machinist, Material Review Board Officers*[4]*, Manufacturing and Instrumentation Engineers*. As the CPSS will change the workflow, we first analysed the types of interactions and interdependencies between inspectors and other workers.

### I.1.1. Integration Points with Designers

In order to provide actionable design patterns to designers/researchers, we have established integration points (Figure 3). We first determined the technical steps that each member of the design team is conducting. We reviewed SARA's research proposal and project follow-up charts to identify key designers to interact with, their milestones and goals (either global or specific). Next, we identified the fieldwork data they might need for their design objectives and when they would need it (timing criteria). This guided us in determining the units of analysis for fieldwork inquiries, as we focused on specific samples of inspector activity (e.g., sense-making and decision-making phases).

---

[4] Committee of experts (usually former inspectors or quality technicians) who treat components labelled as non-conforming in the plant. For example, when the condition of a part is found to be unacceptable, MRB officers take over.

Two Elicitation Interviews (EI) were conducted to elicit designers' needs in terms of data/input. Finally, two additional EI were required to pinpoint how to present and transfer fieldwork data in an exploitable format for designers/researchers. This entire methodological part results in Integration Points that is distributed throughout the chronology of the SARA project (Figure 3).

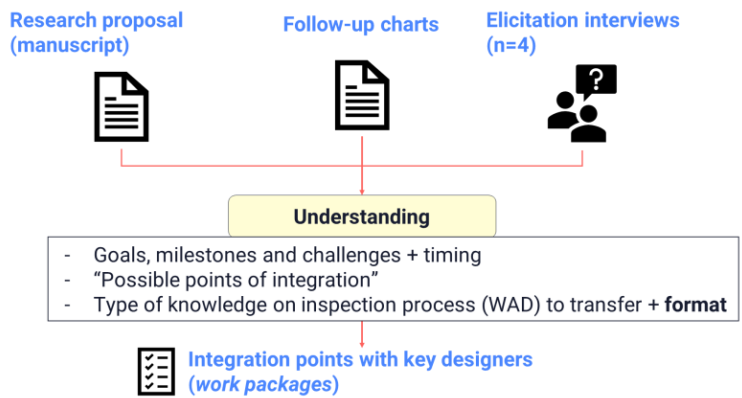

*Figure 3 - Methodology used to provide actionable recommendations to designers/engineers*

## I.2. Processing, Analyzing and Interpreting Fieldwork Data

In this section, we present the methodological steps used to consolidate the raw data obtained through the fieldwork inquiries. During our observations, *Think-aloud* protocols or experiments, we recorded and segmented in real time the fieldwork data collected in an Event Log[5]. Each Log was then analyzed to assign a goal, a task and the underlying cognitive activity to each Event (Table 3). This first step of data processing raised the need to conduct self-confrontation interviews (Section I.1.1) in order to understand or confirm unclear points with inspectors (e.g., knowledge applied, constraints, strategies, information intake mechanisms).

*Table 3 – Example of a computerized Event Log with each feature*

| Time | Event | Associated Task | Goal and Strategies | Cognition (perception, decision, reasoning, problem-solving, errors…) |
|---|---|---|---|---|
| 7:15 | Systemic scan of the upper area of the part | Focus inspection | Rotation of the part under the spotlights to vary angles | |

---

[5] *Record of observable events (actions, verbalization, communication with other people, incidents, changes in the work environment or organization) as they occur, with times noted* and associated to the corresponding task (St-Vincent et al., 2014)

| 7:16 | Stops gaze movement and touches a specific point | Focus Inspection | Tactile sense to confirmation visual cue | Close eye-object distance (increased attention) |
|---|---|---|---|---|
| 7:18 | Defect detection on Tip area (+ verbalization to analyst) | Defect detection | | Inspector: "There's an impact here. You can feel it's sharp around. It's probably a nick" |
| 7:19 | Collects magnification glass and examine the mark with it (+verbalization to analyst) | Interpretation (sense-making) | Use additional material (magnification glass, flashlight) to confirm detection and defect characterization | Inspector: "The impact raised the material on the contour [perceptive cues]. It's typical of the nick." |
| … | … | … | … | … |

Then, we conducted a thematic analysis[6] to cluster the fieldwork data captured from interviews, observations and experiments. This method categorizes relevant data together to establish links among events, phenomena, effects and variables. For example, defect measurement with worn tools (events/constraint) leads to a form of judgment under uncertainty (phenomenon) which could result in the part being sent to the MRB to obtain accurate measurement (effect), especially in critical areas of the part (variables). Thereby, the collection and interpretation of data follows an iterative path to ensure a systemic understanding of the industrial inspection (i.e., identify how the contextual factors of the work domain are shaping inspectors' work activity). Fuzzy points or multifaceted concepts require a confirmation loop that triggers the need for additional *in-situ* data collection and analysis. We finally completed the description and validation of the data incrementally, using semi-directed interviews with stakeholders and data triangulation[7]. The different categories of the thematic analysis are as follows:

- Tasks / sub-tasks: operations carried out by operators to produce a certain outcome
- Purpose: the goal(s) behind each action/intention
- Work organization: aspects concerning the allocation of tasks and jobs among workers, work settings and associated requirements, synchronization with other operations
- Constraints: contextual factors modulating work performance and realization
- Communication: type and format of information that inspectors share or receive from other workers (either verbally or mediated by computers), mutual aid among inspectors
- Gap between WAI and WAD: discrepancies between Work-as-Imagined by the procedures and the actual compensation made by the inspectors to meet taskwork requirements (Work-As-Done)
- Inspectors' cognition:
    - Perceptive cues (either visual, tactile or perceptual-motor): signal or state that triggers the attention while sentencing a defect (T. L. Johnson et al., 2019)
    - Decision-making variables and reasoning: information, criteria and associated cues that are processed in the decision-making process

---

[6] Work Activity Affinity Diagram: Hierarchical classification and grouping of data according to their similarities and how they shed light on user activity and their environment (*The UX Book*, 2012)
[7] Triangulation involves the use of various methods to collect fieldwork data to increase the validity of the results.

- Strategies and interpretation: strategies adopted by expert operators to overcome taskwork limitations. How are inspectors interpreting, using and overcoming procedures limitations?
- Distributed cognition and knowledge: Information flow. What information do inspectors use to sentence a defect? How is this information disseminated in their cognitive environment? Where did they find it and in which format? What are the operational rules (*rules of thumb*) that experienced inspectors follow?
- Alternative solution: alternative workflow that can occur for the same task
- Tacit knowledge: *know-how* and intuitive knowledge that arises with experience

## I.3. Empirical Modeling and Conceptualization

In this article, we define modelling as an approach to disaggregate WAD into several components. As opposed to models - in the strict sense of the term - which are generally grounded in a theoretical/conceptual framework to explain or predict a phenomenon (Waterson et al., 2017). Here, both terms refer to an objective of work disaggregation. We derived data transcribed from fieldwork inquiries into empirical models. We then chose modeling formalisms based on integration requirements and relevant fieldwork data for the system (Section IV.1.2). The collection and modeling of fieldwork data are interrelated: questions raised by designers, unclear operational knowledge or validation of empirical models with SMEs generated multiple iterations between data collection and modeling (Figure 4). This iterative process enables us to develop design patterns that align with the system engineering cycle (Figure 4).

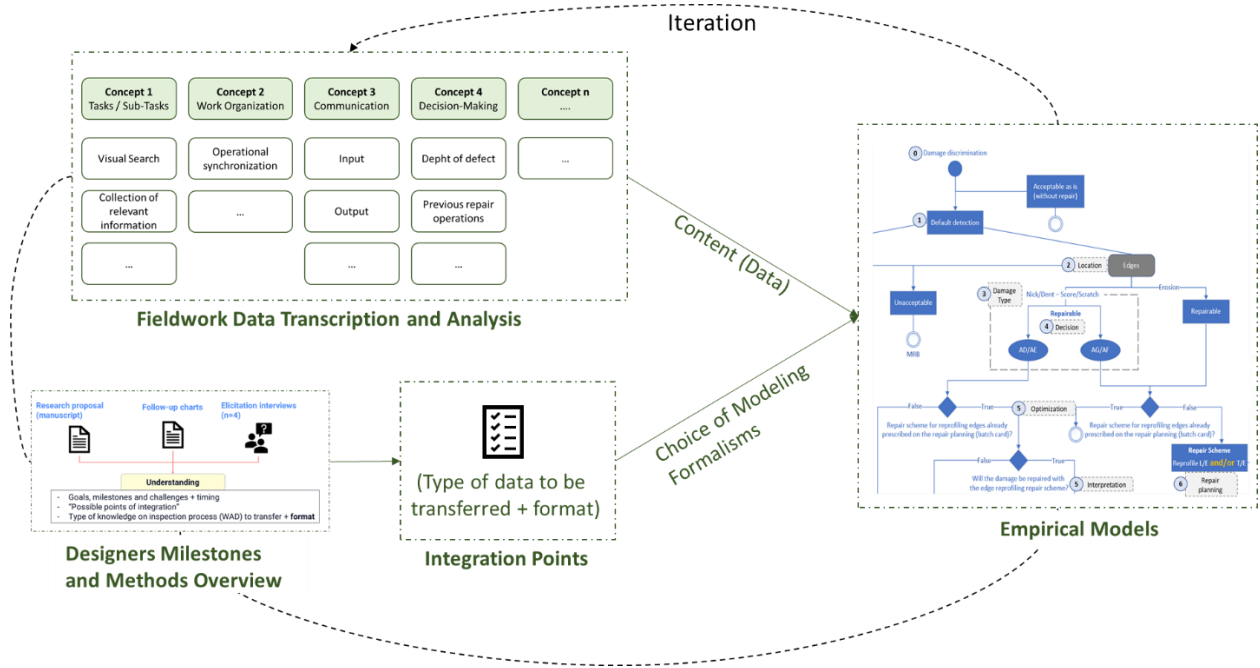

*Figure 4 - Summary of methodological foundations for empirical modelling (actionable design pattern)*

# V. Results

The following section is organized as follows. We first introduce the overall activity of inspection through the general tasks process (workflow). Second, we situate the needs of designers within the context of the inspection workflow. Third, the empirical models developed with fieldwork data are presented. Each subsection corresponds to a specific model, either descriptive or formative. They are depicted with the following points: structure of the models, objectives and exhibited knowledge on industrial inspection. A final section aligns the empirical models with the primary conceptualization of SARA and presents several design assumptions.

## V.1. Industrial Inspection (Sentencing) As-Is-Done on One Part

Sentencing a part involves the combined use of perceptive senses (visual, tactile and perceptual-motor), resources (human, tools, documentation) and cognitive functions (sense-making, decision-making). This activity can be broken down into 5 stages (Figure 5). First, inspectors bring the part(s) to the workstation and collect relevant inspection aids (work instructions, documentation, tools) to start the sentencing (Figure 5 – I. Work Preparation).

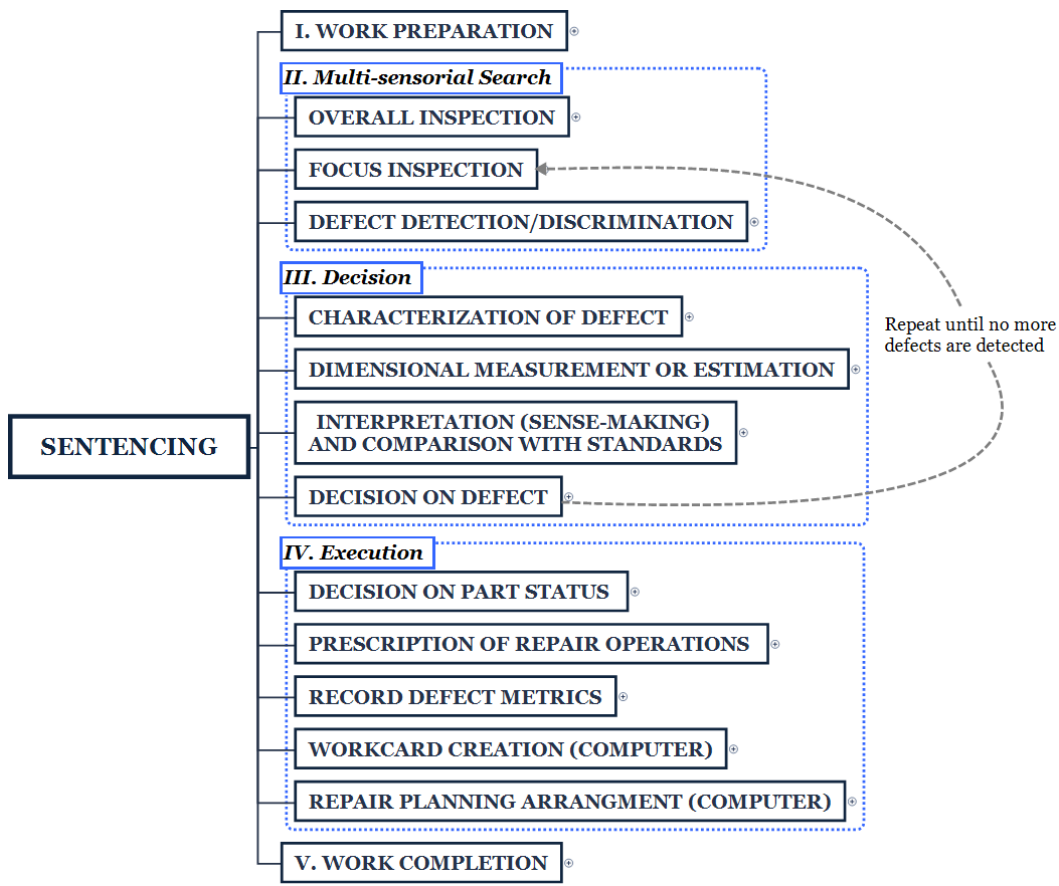

*Figure 5 - Overall sequence of industrial inspection (sentencing)*

Next, they visually inspect the overall appearance of the part to reject any obvious detrimental conditions and then perform a multi-sensory analysis on each area until they detect any damage. During this task, they also verify if previous repair operations were applied as it may restrain further repair scheme applications (decision variables) (e.g., check usual repaired areas for polishing marks) (Figure 5 – II. Multi-sensorial Search). Thirdly, they enter into a complex decision-making loop to diagnose the condition of the defect/part. They characterize the attributes of the defect in order to classify it (Figure 5 – III. Decision). They then measure or estimate the depth of the defect in comparison to the tolerance threshold specified in the standards. Next, they diagnose whether the defect is acceptable (as is), repairable or unacceptable which leads to part rejection. To do this, they process (sense-making) all the decision variables, interpret the validity of previous measures/estimations and eventually confirm their choice with their peers (inspectors, polishers, engineers). The remaining tasks require taking the necessary actions following diagnosis (Figure 5 – IV. Execution): deciding about the overall part's condition (serviceable or unserviceable), prescribing and planning the sequence of repair operations (polishing, machining) and carrying out computerized procedures. Their work is completed when the appropriate information is transmitted to production managing stakeholders and the part (or set of parts) is stored in the corresponding racks (Figure 5 – V. Work Completion).

## V.2. Eliciting Engineering Design and Designers' Needs

Based on fieldwork data collection, interviews with designers and the identification of project designer milestones, the relevant data to be elicited is part of the fourth sentencing steps (out of five) (Table 4, Figure 5). Designers either explicitly mentioned information about their needs concerning fieldwork data during Elicitation Interviews (Table 4) or these requirements have been deduced according to engineering design milestones to be achieved. This dual approach allowed us to identify that half of the relevant data is in the diagnostic loop: key attributes for defect characterisation; processing of information and decision variables; application of situated knowledge and rules according to the constraints imposed by the work domain (Table 4, Figure 5 - III. Decision phase). The overall sequence of sentencing formalized in Figure 5 generates a situated and common representation of inspection tasks. On the basis of this representation and design needs, the relevant types of knowledge found in each task are presented in Figure 6.

*Table 4 - Explicit information (needs) requested by designers*

| Explicit Design Needs | Verbatim |
|---|---|
| Decision Rules | *"Knowing how the inspector makes a decision, we don't have that kind of information (…) We're used to snippets of fuzzy data over a long period of time (...) Scheme or Diagram for us is just wow."* – Automation Manager |

|  | *"It actually just clarifies the process. Right now, we imagine he's doing it* [the inspector] *in a certain way (…) The people who are working here on the inspection* [system]*, they need to better understand the rules and see them in practice."* – Automation Engineer #1 |
|---|---|
| Task Clarification (Defect Characterization) | *"How are they detecting and identifying edges defect? Sometimes, it's not perceptible in our images' acquisition"* – Automation Engineer #2 |
| Fuzzy Point | *"Sometimes it's not clear in the Engine Manual* [SOP]*. They are no criteria for every class of defects. What are they doing* [inspectors] *in cases like that? Which tolerance should be applied? (…) If we have this information, we'll code it."* – Automation Engineer #3 |
| Overall View of Fieldwork Sequence | *It will help us to move forward* [fieldwork data]*. Unfortunately, we rarely have the opportunity to do this work* [fieldwork inquiries]*. Our job is much more how to inspect a lot of parts and when we have questions, we ask them* [clients] *(...) Then, we try to put the puzzle back together."* - Automation Manager |

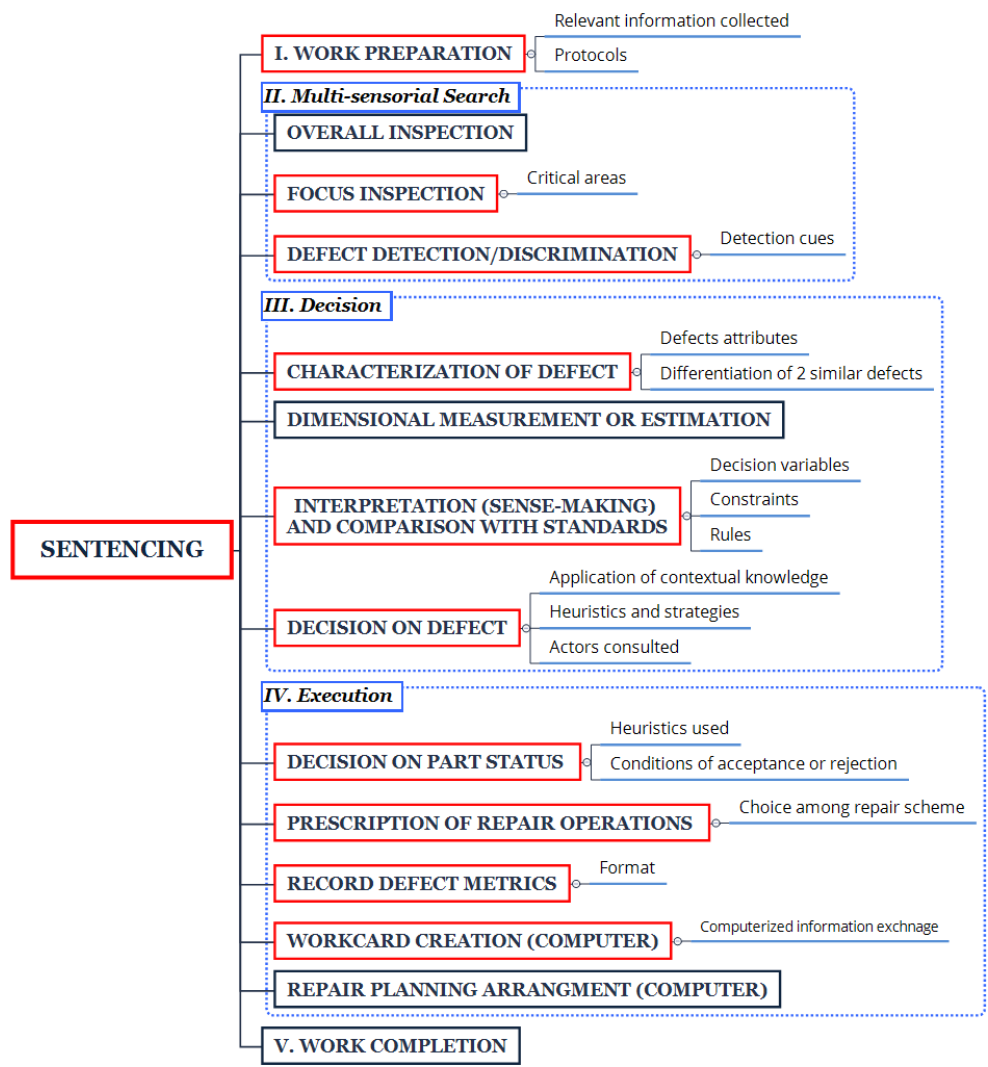

*Figure 6 - Relevant tasks (in red) and data of the sentencing process to be obtained and modelled according to design needs*

The following subsections present the translation of the fieldwork data collected into empirical models that fulfill two functions: 1) a thorough multi-layer description of the complex socio-cognitive activity of inspectors 2) an abstraction/disaggregation of fieldwork data into actionable design patterns that support and guide the innovation process.

## V.3. Using Models to Formalize the Practice of Industrial Inspection and Pre-conceptualize a Cyber-Physical-Social System

### V.3.1. Descriptive/Formative: Activity Model (Work-As-Done)

The Activity Model is a systemic hierarchical disaggregation of work that formalizes tasks, operations, resources consumed (information, knowledge, protocol), the decision-making flow and modulating factors (Figure 7, 8). It is an extended version of the overall sentencing sequence that emphasizes the situational aspect of work (i.e., how is the work done in a real context and what are the decision variables that inspectors must deal with?).

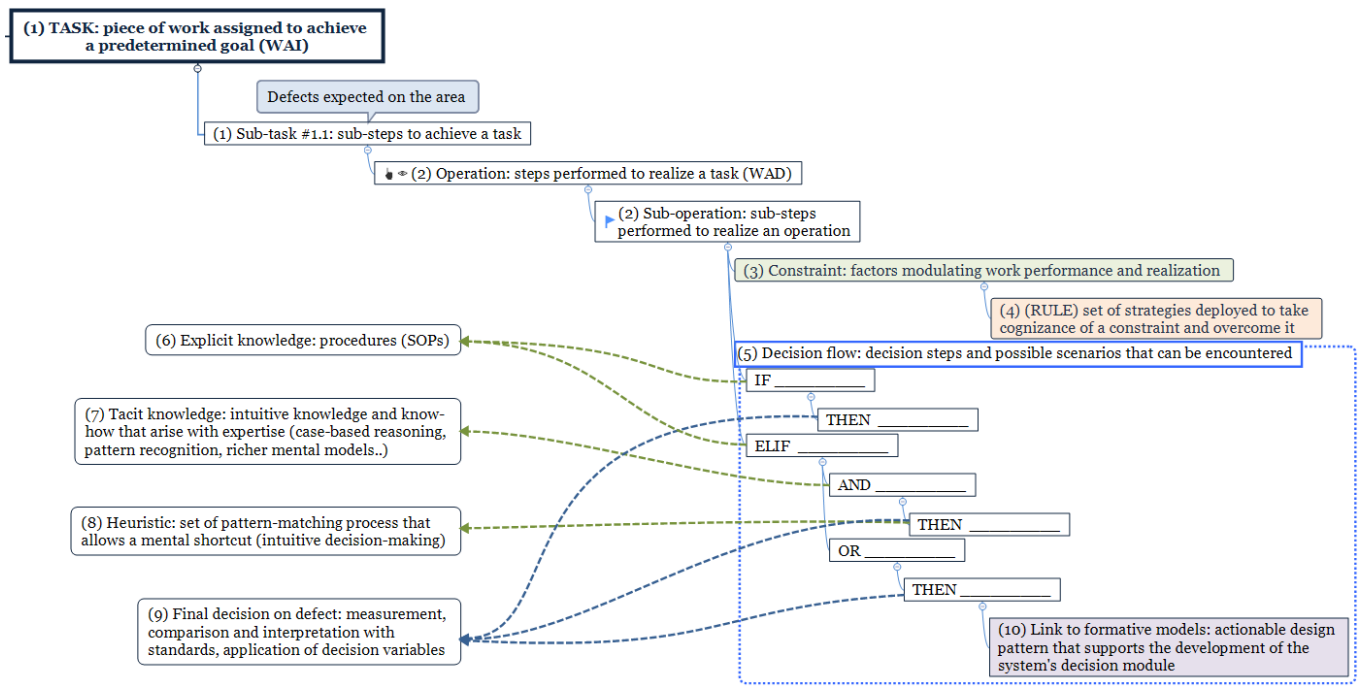

*Figure 7 - Activity Model Structure and feature explanations*

We call variables the "decision arguments" that shape the inspectors' decision-making process, whether they be economic, socio-organizational, derived from safety policies/procedures, and/or of course, the tacit knowledge developed with experience. The modeling formalism emphasizes the socio-cognitive dimensions of inspectors' work; however, the perceptual-motor and physical and social dimensions are also integrated (Figure 8).

Inspectors' activity is represented in a systemic way by showing how work domain constraints shape the inspectors' cognition and action. The institution prescribes sets of rules and this model shows how they are processed (sense-making) and how knowledge and variables are applied in regards to domain constraints (decision-making). Each step of the decision-making flow is presented with alternative courses of action (different possible scenarios when encountering a defect) and the knowledge mobilized whether procedural or tacit (Figure 7). As some constraints are very situated, the model path highlights the strategies deployed by inspectors to overcome the constraints and procedures limitations (Figure 8).

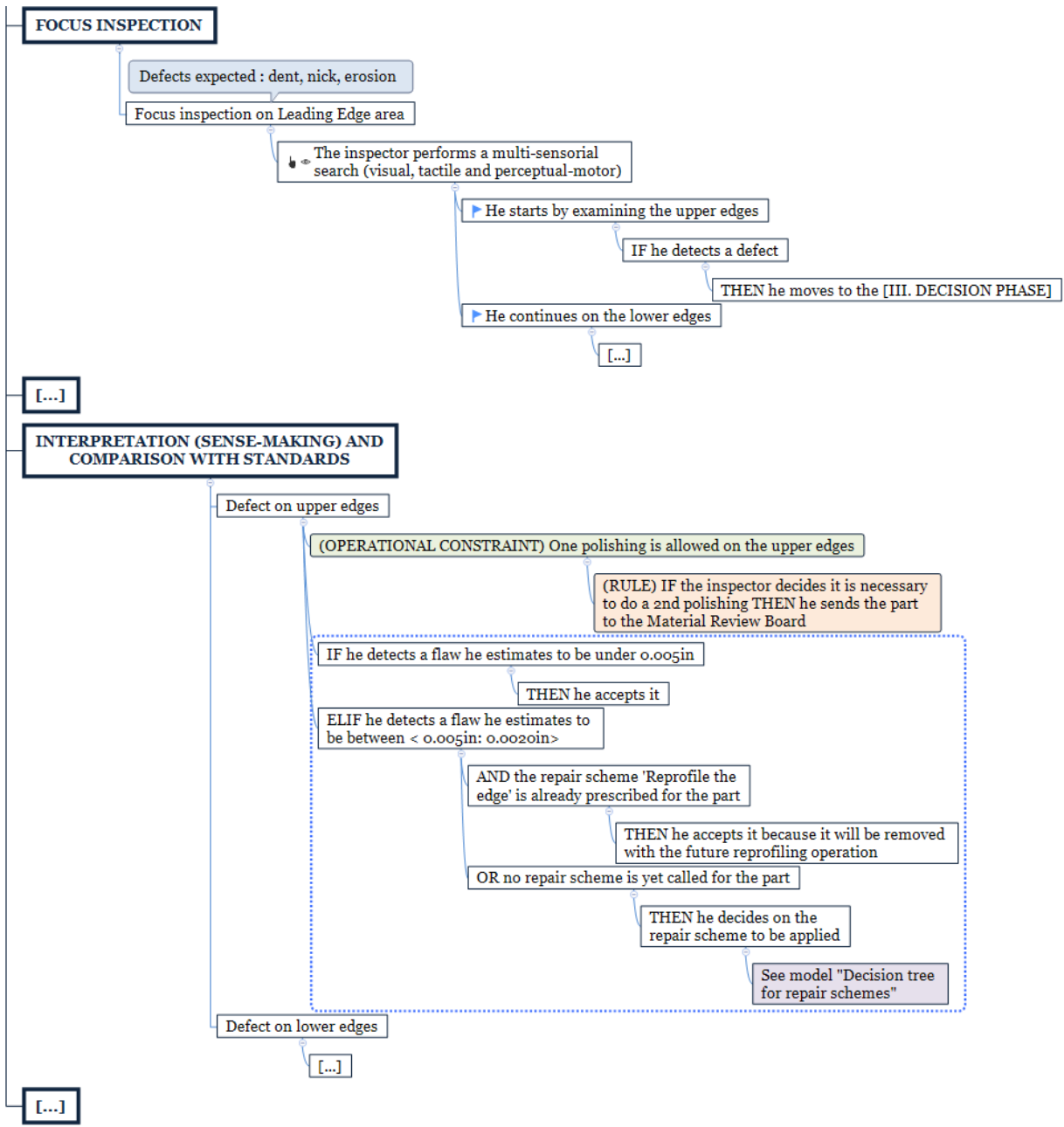

*Figure 8 - Activity Model Example (screenshot)*

The main decision variable relates to the physical characteristics of the damage. SOPs' present the tolerance threshold for each type of damage and its location on the component. However, inspectors consider other variables to decide whether a defect must be repaired. A list of the other types of decision variables used by inspectors is presented in Section V.3.5.

Some repair operations can only be applied a certain number of times and, in case the repair operation has to be applied once more, they rely on operational rules, heuristics or strategies to overcome those limitations. Knowledge of the repair process allows them to know whether certain defects could be removed by subsequent machining operations (Figure 8). Therefore, they assess in real-time the need to prescribe an additional repair operation.

The following two sections present formative models that align with design needs and focus on specific sentencing tasks (V3.2, V3.3). The models attempt to answer the following question: what knowledge should be included in the system database to perform the targeted sentencing tasks?

### V.3.2. Formative: Defect Characterization with Fuzzy Criteria

The Defect Characterization Model defines the attributes that characterize each defect with fuzzy criteria: physical appearance, attributes/conditions and expected (problem) areas on the component (Figure 9). Identifying the type of damage is mandatory as tolerance thresholds differ according to the type of defect. For the same depth and width, one defect could be repairable and another irreparable. Inspectors have developed strategies over time to better recognize defects. This model details the procedural and tacit knowledge mobilized by inspectors to classify damage types (first step of the decision-making phase).

We built the model on the basis of internal documentation and *in situ* analysis. We obtained valuable results from *think-aloud* protocols where inspectors verbalized the attributes that characterize one defect or differentiate two relatively similar defects (e.g. nick and dent, Figure 9). These attributes are shown in blue in Figure 9.

For example, a "nick" is an impact defect that causes a vertical movement of the material along the defect contour (raise material). The presence of high material is distinctive of nicks. Inspectors rely on tactile or perceptual-motor feeling (using a stylus) to detect any material variation on the contour or on the bottom (*floor*) of the defect (Figure 9). A "dent" has similar properties but with a horizontal material displacement. Finding high material around the damage is unusual. Moreover, inspectors have developed the ability to distinguish in-flight defects (external) and handling damages (internal). Based on the type of defect and its location on a part, and on the inspector's overall knowledge of the engine (rich mental model), inspectors rely on causal reasoning to produce hypotheses about the cause of the defect. In doing so, they can alert manufacturing engineers if they suspect a faulty internal process that generates handling damages.

| Nick | Dent |
|---|---|
| **Appearance** Light cavity (asymmetrical - heterogenous) | **Appearance** Strong cavity (uniform - homogeneous) |
| **Attributes / Conditions**<br>- Tangential or angular impact<br>- Uniformity = no<br>- Defect 'floor' = sharp (stylus)<br>-Defect edges = sharp (touch= 'catchy')<br>-Highmaterial = yes (usually)<br>-Material displacement = vertical (raise material)<br>- Material removal = maybe | **Attributes / Conditions**<br>- Tangential or angular impact<br>- Uniform = yes<br>- Defect 'floor' = 'rounded' (stylus)<br>- Defect edges = smooth (touch= 'soft')<br>-Highmaterial = maybe (unusal)<br>-Material displacement = horizontal (material compression)<br>- Material removal = no |
| **Expectations**<br>- Concave airfoil (near L/E) | **Expectations**<br>- Concave airfoil<br>- Edges |
| **Fan Blades**<br>- Same tolerance limits in less critical areas (AE, AD) but deviations are observeable in critical areas (AG, AF, Root)<br>- Inspectors detects high material around defects using a scriber (stylus) ||

*Figure 9 - Defect Characterization Model (screenshot). The original diagram contains the 16 defects that could appear on the concerned component*

### V.3.3. Formative: Decision Tree Model for Repair Operation Prescription

The Decision Tree Model is a sequential diagram that formalizes how inspectors decide which of the different repair schemes to prescribe. The model is structured on the basis of the inspector reasoning, starting with defect detection (Figure 10):

1) Detection (discrimination) of defects: A multitude of defects can be concentrated in a single area. When a defect is judged to be borderline within a tolerance threshold or when a salient feature (unusual expectation) is spotted by the inspector, this operation stops and the decision-making process starts.
2) Location of the defect on the concerned part: they locate the area in which the defect appears. Tolerance thresholds differ from zone to zone. Criteria are more restrictive in critical areas.
3) Classify damage type: crucial as tolerance thresholds change from one defect to another
4) Dimensional measures (this step doesn't appear on the UML formative model): physical measurement of the defect in terms of depth, length and width with metrological tools.
5) Decision: they decide whether the damage is acceptable (as is), repairable (requires various repair operations to remove it) or unserviceable (the component is sent to the MRB)
6) Optimization/interpretation: inspectors evaluate whether a defect could be removed by an operation that has already been prescribed in the repair planning[8] (optimization). They assess in real-time whether a repair operation already included in the repair planning could remove the defect before prescribing another. By doing so, they improve process efficiency and components lifecycle (each polishing removes parent material which reduces the lifetime of a component). "*Is there any operation in the repair planning that could remove this defect or should I prescribe another operation?*" Inspectors carry out the same type of interpretation/optimization loop after verifying the dimensional measures.

---
[8] Repair planning: sequence of repair operations prescribed for one part

7) Repair planning update: (re)plan the sequence of operations required to remove the defect and improve the surface finish.

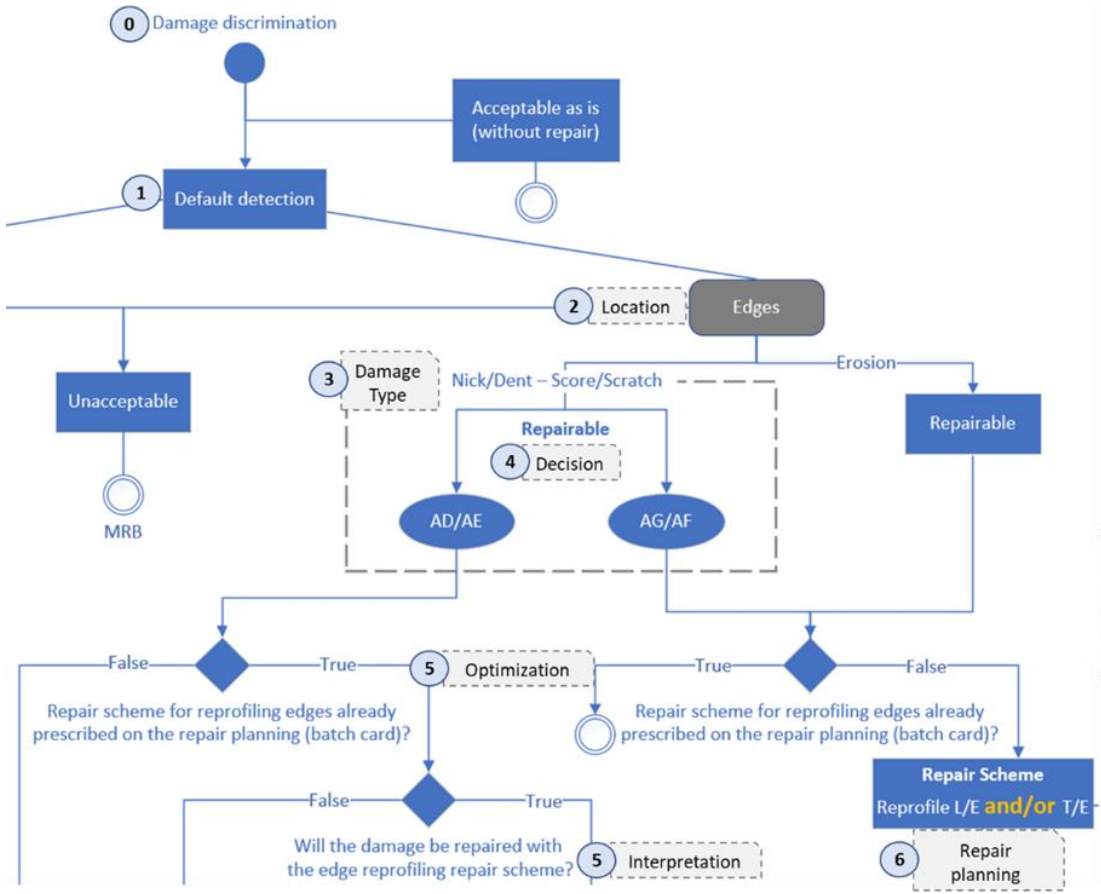

*Figure 10 - Computerized version of one part of the decision tree (screenshot). Due to confidentiality concerns, we have deleted sensitive information.*

### V.3.4. Descriptive: Distributed Cognition and Information Flow Model

The Distributed Cognition and Information Flow Model is a functional diagram that shows how information is structured and propagated in the work system. The model is organized around the 5 main inspection phases. It details the tasks for which inspectors use external information resources: other actors of the socio-technical system, standard procedures or job aids. Experiential (tacit) knowledge acquired through experience is also incorporated into the model (internal information resource stored in long-term memory).

Inspectors are interconnected with polishers, machinists, MRB officers and production managers at different steps of the sentencing sequence. They are in constant interaction with the operators and have developed collective strategies to assess repair possibilities ("is it possible to polish this area without exceeding the limits of material removal?") or to facilitate each other's work by transmitting additional instructions. Some repair operations automatically restore the whole part. One operation consists of reprofiling both side of a part. When a side is acceptable as is, inspectors add additional instructions on the computerized work card specifying that only the affected part needs to be reprofiled.

*Table 5 - Caption of the Distributed Cognition Model*

| Features | Illustration |
|---|---|
| Task | Overall inspection |
| Informational resource (SOP, protocol, internal documents) | Minor Damage Record Sheet (HRS3439) |
| Data and type of data | Airworthiness (cycle flight and time) — Integer |
| Operational expertise (strategies, tacit knowledge, heuristics) | 🧠⚙ |
| Actor(s) involved | 👷 |
| Condition / decision variables | Relevant information collected |
| Sequential flow | ⟶ |
| Action of the sequential flow | — communicates ⟶ |
| Information intake | ∘─ ─ ─ ─ ─ ─▷ |
| Association | · · · · · · · · · · · · · · · |

This form of collective cognition prevents unnecessary repairs and thus enhances turnaround time and part lifecycle. As the sentencing output determines the various operations that a component must undergo to be restored, inspectors are also "work specifier". They are responsible for updating the progression on the production board and informing the production managers each time a set is completed or a change is made or required on the repair schedule (repair order, adding or deleting repair operations). With this information, planners can evaluate process capabilities and redistribute resources efficiently.

If an inspector measures or estimates a defect exceeds the limits or if he does not have sufficient information to make a decision, they send a computerized report to the MRB detailing the parameters of the defect (physical characteristics, location on part), the causes of rejection and the variables that determined their choices (exceeding acceptance limits, presence of previous repair operations, location not defined by the standards). MRB officers then take over, carry out background investigations (i.e., the track record of the part), contact the Original Equipment Manufacturer (OEM) and ask for laboratory analyses to give the inspector new instructions. As MRB procedures are costly, time-consuming and risk bottlenecking the process (Figure 11), inspectors have developed strategies over time: They can help each other by exchanging information about past decisions made on similar cases (collective case-based reasoning) or perform a minor repair with sandpaper. Also, as MRB workstations are located on the shop floor, they are often directly consulted by inspectors without the need to create a report.

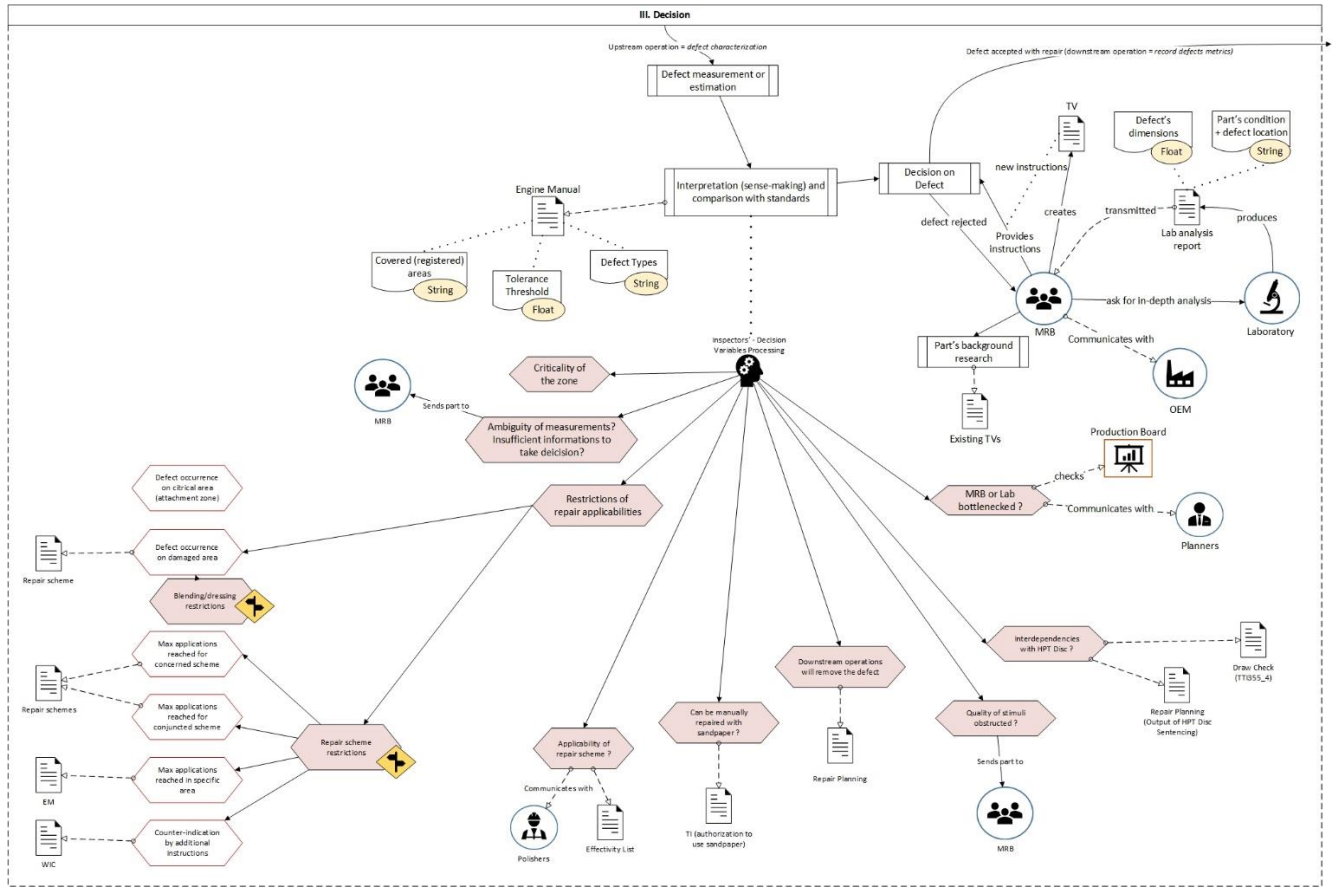

*Figure 11 - Distributed Cognition and Information Flow Model (screenshot)*

### V.3.5. Pre-Conceptualization of SARA Using Empirical Models

This subsection summarizes and exemplifies how the empirical models support and guide the pre-conceptualization of SARA. We first present the design assumptions behind each model: how they can be used as projection tools for designing the system and conceptualizing Human-Agent Teamwork. Then, we present an abstraction of the different domain decision-making variables that arise from the empirical models and that must be reallocated to the future Human-Agent Team.

**Activity Model.** Task-work requirements, contextual factors and operational demands are formalized in this model. All tasks and decision variables are made explicit and those that must remain with the future Human-Agent Team are identified and classified (Table 6). Based on technical capabilities and task-work requirements, the design team, early in the engineering cycle, can identify the decision variables and tasks that will be reallocated/redesigned to the system and the operators (Appendix 1). In doing so, the conceptualization of the future Concept of Operations (ConOps) will articulate technical and ecological validity, which ultimately favor system fit with workers' practices (Appendix 1). Finally, this model is a reference for the future development of more dynamic models that simulate and evaluate the validity of several ConOps. Actionable design patterns can be abstracted from the decision-making flow and translated into machine-readable components. The structure used (IF, ELIF, AND, THEN) can facilitate this translation process. Finally, this model provides a reference for the future development of more dynamic models that simulate and evaluate the validity of several ConOps.

**Defect Characterization Model.** The system should be able to define the relevant combination of attributes that categorize each defect. This model provides relevant cues that can either guide or support system development. Our preliminary results suggest that the Defect Characterization Model supported researchers evaluating the validity of computer vision approaches to describe defects (i.e., to extract relevant information from a 3D point cloud of a defect). Indeed, it improves defect characterization approaches and supports the exploration of alternative possibilities to classify several types of defects (Appendix 1). Other applications are conceivable with machine learning methods (e.g., calibration of neural networks weights using fieldwork knowledge).

**Repair Prescription Decision Tree Model.** This model provides guidance for the decision and execution module of the CPS. UML-based formalism is easily translatable into program development language (Appendix 1). Also, decision trees are one of the most interpretable machine learning methods. Since explanation is a mainstay of safe AI, this model can be used as a reliable structure to generate explainable outcomes.

**Distributed Cognition and Information Flow Model.** Inspectors' cognition is distributed in a decision-making ecosystem composed of several actors involved in the reparation process. Integration of interdependencies between actors when designing the system reduces the risk of unforeseen and negative operational impacts. In order to ensure the ecological validity of SARA, the same data should be transferred to the MRB when assessing a defect that exceeds the limits (Appendix 1). Visualization of the information flow and interrelation between stakeholders can be used as input for the data management structure (Appendix 1). Finally, this model could support the calibration of future agent-based simulations for evaluating the system.

**Decision-making Variables.** The inspectors' decision-making process is complex, dynamic and modulated by contextual factors and operational constraints. Indeed, they process several variables that can either be specified in inspection aids documents, arise from operational constraints or from experience (e.g., case-based reasoning). These variables are related to the physical characteristics of the defect, the overall condition of the part and its history (e.g., previous polishing area) or the available course of actions (Table 6). From an automation perspective, some of the variables are likely to disappear, especially the ones associated with human dimensional measurements (Table 6 - #4). However, the engineering team must assess system capabilities in taking cognizance of all these variables in order to ensure the validity of the outputs. From a Human-Agent Teamwork perspective, these variables need to be reassigned taking into account several factors such as human/agent capabilities and interaction requirements.

*Table 6 - Formalization of sentencing decision variables*

| # | Decision-Making Variables | Likely to Disappear | Future Human-Agent Teamwork Projection |
|---|---|---|---|
| 1 | Depth, width, length, circumference of a defect | No | Essential. Defect's physical characteristics |
| 2.1<br>2.2 | Previous repair on a damaged area;<br>or on a restricted location | No | Essential. Restrictions due to material thickness or due to attachment of the area with an adjacent component |
| 3.1 | An already prescribed repair operation will eliminate the defect | No | Essential heuristic to optimize a part's lifetime and process turnaround time |
| 3.2 | Applicability of the repair operation: confined space, maximum application of repair operations reached, etc… | No | SARA could struggle to apply repair on the same area as inspectors |
| 4 | Ambiguity of dimensional measures and estimations (e.g., worn gauge) | Yes | SARA will carry out dimensional measurements |
| 5 | Minor repair (using sandpaper) could remove the defect | Yes | More repair operations are to be expected with SARA |
| n | … | … | … |

## VI. Discussion & Conclusion

Designing a Cyber-Physical-Social System (CPSS) in complex manufacturing environments requires a shared understanding of the operating context among designers. How do operators make decisions? What are the criteria (variables, rules, etc.) and constraints within the socio-technical system that shape their cognition and action? Such an understanding informs designers about work complexity (Roth et al., 2019). However, fieldwork data transmitted to design engineers is seldom actionable (Emmenegger & Norman, 2019; M. Johnson et al., 2014). Further steps are required to translate the data into actionable input. This paper takes a step in this direction by providing a method for formalizing relevant fieldwork data that aligns with system design milestones on a concrete industrial case study.

The section is organized as follows. We first discuss the methodological set deployed: from fieldwork data collection to modeling formalisms. Next, we examine the impact of empirical models i) to support the design of CPSS in a real-world setting and ii) to provide support and guidance to multidisciplinary design stakeholders. Finally, we present the limitations of the current method and some directions for further research.

### VI.1. Discussion and Lessons Learned

**Fieldwork Research and Work-Centered Approach.** The combination of data collection methods deployed in this research has generated a multi-layered description of industrial inspection: from a socio-technical point of view to a thorough study of inspectors' cognitive activity. Indeed, empirical models support a micro, meso and macro perspective of "work". They represent the physical, cognitive and social dimensions of the actions taken to produce correct decisions (i.e., Work-As-Done and Work-As-Imagined). Aircraft engine inspection is part of a complex and highly regulated socio-technical system where several actors ensure compliance with regulations, rules and protocols. We have identified the semantics of the problem space through a systemic approach to work situations (or work-centered approach): what are the work domain constraints and how they are shaping the cognition and action of inspectors (RQ1). The Distributed Cognition Model abstracted the interdependencies and shared strategies between inspectors and the other actors. First, it highlighted how inspectors are embedded in a decision-making ecosystem. Second, the model supported the abstraction of initial assumptions concerning the potential impact of automation on the information flow, interdependencies between actors and function allocation (at least, on the three parts selected for the project). Then, studying the cognition of experts' through *think-aloud* protocols and experiments enabled us to delve in the core of inspectors' work. It specified the knowledge (*know-how*) and variables processed during each step of the sense-making and decision-making process. This "know-how" is not stored in SOP and therefore requires a robust methodology to extract and represent it from the heads of SMEs. Indeed, inspectors rely on a substantial amount of tacit knowledge, heuristics and strategies (individual or collective) to process decision variables, produce correct outcomes, save part's lifecycles and reduce turnaround time (Section V.3). Our results are consistent with several studies that suggested that only a bottom-up approach grasps the highly situated constraints, knowledge, motivations and affordances (variables) that experts' deal with in real-work settings (T. L. Johnson et al., 2019; Meinherz & Videira, 2018; Morineau & Flach, 2019).

**Modeling Formalisms.** Although fieldwork data are extremely valuable in conferring ecological validity to the system being designed, field researchers seldom provide exploitable design patterns from a software/automation engineering perspective. Often, data are conveyed to designers in a purely descriptive form i) lacking integrability and guidance to the milestones being pursued by designers ii) and generally misaligned with the overall system design objectives. One needs to find a balance between descriptive and formative (or normative) output. Indeed, descriptive models are invaluable in disaggregating the characteristics of WAD within the embedded socio-technical system: tasks, work requirements, domain constraints, actors' role, skills required, knowledge applied, decision flow and sense-making process. This knowledge informs the design stakeholders of work complexity and situates the process tasks that are subject to automation. However, descriptive models are seldom actionable. Bridging the gap between descriptive and formative output involves aligning the fieldwork research to the design objectives. In our methodology, we identified the design milestones, the actors involved and their needs, which ultimately guided the data collection (e.g., different units of analysis) and formalization steps (e.g., "how to make fieldwork data exploitable"). Theses steps are by nature iterative. Models are refined in response to new questions from designers (as they progress through the design process), resulting in additional sessions of fieldwork inquiries and modeling (RQ2).

Empirical models are aligned with the maturation of certain characteristics of the system (software modules): defect detection and filtering (Defect Characterization Model), decision algorithms (Activity Model) or execution modules (Repair Prescriptions Decision Tree). Taking into account the operating context, they will also guide the specifications of Human-Agent Teaming (Activity Model and Distributed Cognition Model).

**Preliminary Conceptualization of a Cyber-Physical-Social System.** Formalizing the current workflow with models has highlighted the tasks, variables and knowledge that SARA needs to incorporate/execute. We argue that prior to the implementation of an intelligent system in real-work environments, the specific application of knowledge must be elicited and represented to the design team in a way that can provide in-depth insights and actionable patterns for system design. Since agents will perform a certain amount of mental work, it is crucial to identify not only the knowledge required in the agent database, but also how to apply it, regarding the constraints that shape the work domain and task-work requirements. Empirical models showed all the actions performed by inspectors, whether prescribed (WAI) or self-added (WAD). This first layer of analysis formalized the fundamental tasks that must remain in the future human-agent workflow. One of the main limitations of traditional function allocation methods for Human-Agent Teaming is related to a lack of consideration of work requirements i.e., "what needs to be done" (Roth et al., 2019). To ensure the ecological validity of the designed system, the operating context must be understood and decision choices must be based upon this comprehension. However, the reassignment of work components to an agent will inevitably transform the work system, resulting in the modification, deletion of tasks from the "manual" work content and the emergence of new ones (Hew, 2017; T. L. Johnson et al., 2019).

Therefore, future design steps will explore cyber-physical-human (joint) workflows through scenarios, simulation and tests.

**Interdisciplinarity, the Future of Field Researchers in Complex System Design.** Designing a complex CPSS requires collaboration between several disciplines. To orchestrate it, empirical models endorse a boundary object function that feeds into a design thinking process. They create a shared understanding of work complexity and draw attention on challenging elements in the system design stages. Indeed, a holistic perspective is brought to the current design process, in terms of technical-organizational-social congruence rather than only technological maturity (Cabour et al., 2019). Maturing system's technological components is obviously crucial, however, its implementation, utility, usability and acceptance by end-users depend on additional criteria:

- Accuracy of the analyses carried out: variables used in the decision-making process, correct application of knowledge and concepts, reliable data and algorithms, trust in the output, etc.
- Quality and interpretability of the information transmitted (in line with operational needs and task-work requirements)
- Congruent Concept of Operation: (re)allocation of tasks (or decision-making variables), role and responsibilities among humans and agents that consider operational interdependencies (Figure 12)

By positioning themselves between technology and the socio-technical system, fieldwork researchers can corroborate the right technical-ecological fit in the design of future CPS(S). Several authors mention this point as one of the biggest challenges in implementing new autonomous technologies in real-world settings (Boy, 2020; Cabitza & Zeitoun, 2019; Cutillo et al., 2020).

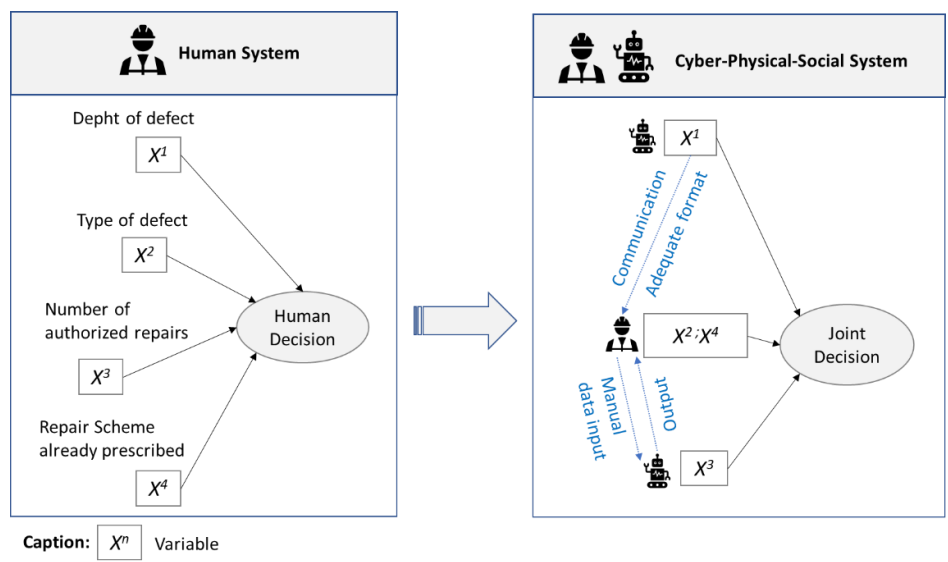

*Figure 12 - Example of jointed decision-making process in a Cyber-Physical-Social Inspection System*

### VI.2. Limitations, Future Work and Conclusion

This paper has some limitations. The process of data collection and validation were time-consuming, reducing the applicability of the methodology to short-term projects. The disparity of the actors observed/interviewed lengthened the time and effort needed to reach a consensus. Some researchers used group interviews to speed up the consensus building which could be of interests for future fieldwork inquiries (Hobballah et al., 2018; Milton, 2007). Moreover, even with the precaution of triangulating qualitative data, there may be a gap between what we collected and the reality of everyday practices. In other words, we have elicited what we observed at *time t* until we reached a point of redundancy (specific to the subjective judgement of the researchers). Second, the success of this method relies on the commitment of experts, which requires an appropriate change management process to allay fears and resistance (e.g., fear of job loss). Third, the method was developed in an existing environment. The output generated through the method is thereby idiosyncratic. These findings should be considered in the context of one case study that selected three engine components from a fleet of over 1,000. Finally, the results presented are overly descriptive and assumptive so far. As we proceed the research project, we will provide concrete cases of the use of empirical models in the engineering cycle and an evaluation of a work-centered approach for the socio-technical maturation of a CPSS. And then, we will conceptualize a generic framework that generalizes the methodological set deployed here.

However, we hold the methodology is context-independent, as long as the essential features are retained: fieldwork inquiries are translated into empirical models that i) align relevant fieldwork data with system design ii) suit the cross-functional integration requirements and raise the challenges pursued by the engineering team iii) enable an early conceptualisation of the future Concept of Operations (ConOps). The units of analysis for fieldwork inquiries and empirical modeling must be adapted according to the work environment, the engineering needs and/or the nature of the system being developed. However, further research should focus on developing methods that use the results of a work-centered approach as input to define the ConOps, as it determines CPSS's productivity-related factors and success of implementation. For this purpose, we began to create future workflow scenarios on the basis of empirical models, technological capabilities and human-agent interaction requirements.

In this paper, we have revealed the potential of a work-centered approach for designing Cyber-Physical-Social System (CPSS) in real-world settings. Our contribution is threefold. First, we offered a multi-layer perspective of industrial inspection with a deep level of granularity. We have delineated how inspectors make decisions in a real, complex and knowledge-rich socio-technical environment. Second, we have detailed an effective methodology for eliciting and translating relevant fieldwork data into exploitable insights. Exploitable insights refer to the specific application of knowledge in the operating context (e.g., rules and regulations to respect). Indeed, the Cyber-Physical System (CPS) being designed must respect the application of these rules to generate reliable output (ecological validity). Descriptive models exhibit the socio-technical complexity of inspection and formative models align the data generated with technical development (software modules, Human-Agent Teamwork). Third, we have established useful empirical modeling formalisms that provide guidance to designers by showing how algorithms should behave in the problem space (operational constraints, institutional rules and procedures). They also formalize the functions that the CPSS is expected to perform. This primary conceptualization will benefit the design of CPSS by best (re)allocating tasks, decision variables, roles and responsibilities between actors and software agents. Application of this type of methodology will be crucial for companies. To be properly designed, implemented and accepted, AI-based technologies must fit into the operating context and the workflow of subject-matter experts.

**Acknowledgments**. We would like to thank all those involved in the SARA project for their support and trust in exploring new ideas. Special thanks are expressed to the Montreal plant operators, managers and engineers who participated in the fieldwork sessions. We would also like to thank the members of the technical design consortium for their helpful advice, feedback and support, including R-A. Cusson (Rolls-Royce Canada), M. Chenard (AV&R), D. Laurendeau (Laval University), S. Yacout (Polytechnique Montreal), S. Achiche (Polytechnique Montreal).

**Funding.** This research is supported by the *Consortium for Research and Innovation in Aerospace in Québec* (CRIAQ), funded by Mitacs accelerate program (contract #*Manu-1712* - IT11797). The findings and conclusions in this report are those of the authors.

**Authors Contribution.** Conceptualization, G.C and E.L.; methodology, G.C.; fieldwork inquiries, G.C.; data analysis, G.C.; empirical modeling, G.C.; validation, G.C.; project coordination, G.C., E.L. and S.B.; supervision, E.L. and S.B.; writing – original draft preparation, G.C.; writing – review, feedback and editing, E.L.; writing – final review, E.L. and S.B.; project administration G.C. and S.B.; and funding acquisition, S.B.

**Conflict of interest.** The authors declare no financial interests in the design and deployment of SARA.

**Abbreviations.** The following abbreviations are used in this manuscript:

ConOps  Concept of Operations

CPS     Cyber-Physical System

CPSS   Cyber-Physical-Social System

HAT    Human-Agent Teaming (Teamwork)

MRB    Material Review Board

MRO    Maintenance, Repair and Overhaul

SARA   *Système d'Analyse et de Réparation Automatisée* (Automated Visual Inspection, Sentencing & Dressing)

SOP     Standard Operating Procedures

SME    Subject-Matter Expert

WAI     Work-As-Imagined

WAD    Work-As-Done

## Appendix 1 – Empirical Models and Design Assumptions*

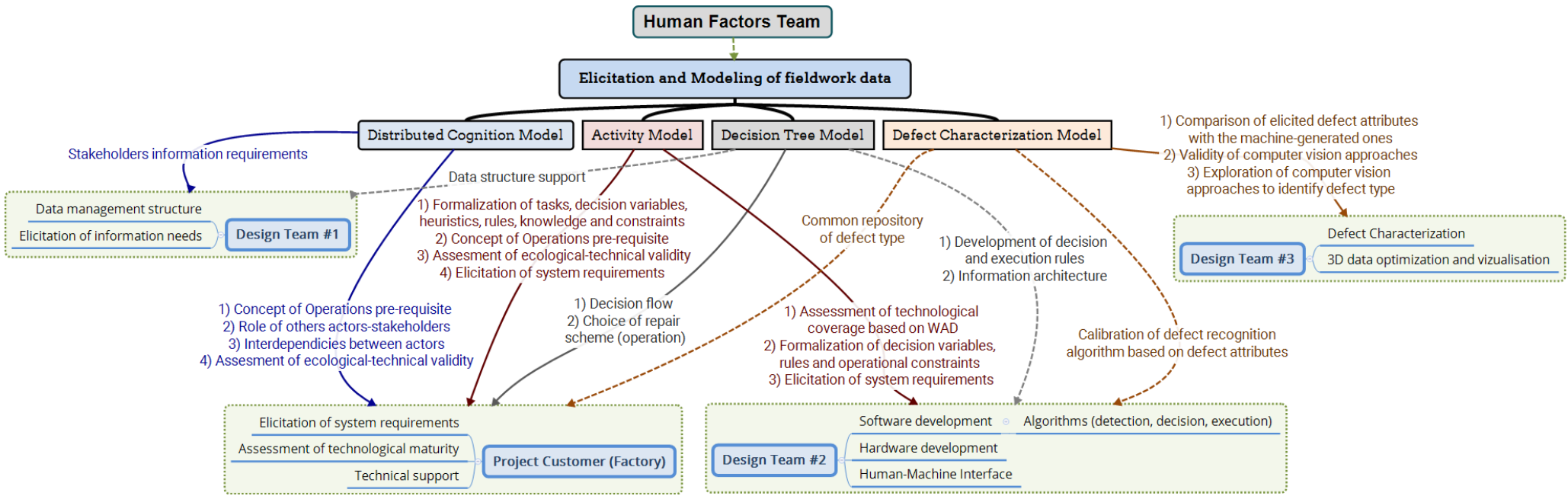

*\* The regular lines represent the actual contribution of the models and the dashed lines the future assumptions.*